\documentclass[aps,pre,groupedaddress,amssymb,amsmath,pra,floatfix,showpacs]{revtex4}
\usepackage{epsfig}

\begin{document}

\title{J-transform applied to the detection of Gravitational Waves:\\
preliminary results}
\author{Daniel Bessis and Luca Perotti}
\affiliation{Department of Physics, Texas Southern University, Houston, Texas 77004 USA}
\date{\today}

\begin{abstract}
We propose to apply to the detection of Gravitational Waves a new method developed for the spectral analysis of
noisy time-series of damped oscillators. From the Pad\'{e} Approximations of the time-series Z-transform, a Jacobi
Matrix (J-Matrix) is constructed.

We show that the J-Matrix has bound states with eigenvalues \textit{strictly
inside} the unit circle. Each bound state can be identified with one precise
damped oscillator. Beside these bound states, there is an essential spectrum
sitting on the unit circle which represents the noise.

In this picture, signal and noise are clearly separated and identified
\textit{in the complex plane}. Furthermore, we show that the J-transform
enjoys the \textit{exceptional feature of lossless undersampling.}

We take advantage of the above properties of the J-transform to develop a procedure for the search of Gravitational
Wave bursts in interferometric data series such as those of LIGO and VIRGO projects. Successful application of our
procedure to simulated data having a poor signal to noise ratio, highlights the power of our method.

\end{abstract}

\pacs{07.05.Kf, 95.75.Pq, 95.55.Ym}
\maketitle

\newpage

\section{Introduction}

Experimental time-series are always affected by the presence of noise. As long as the signal to noise ratio is not
too poor, several filters are available to denoise the data within the framework of Fourier analysis and its
variants. All such techniques, on the other hand, have drawbacks \cite{jpa} and fail when the signal to noise ratio
approaches $1$. It is exactly this very high noise case researchers have to deal with when analyzing LIGO and VIRGO
data. The preliminary evidence we present here shows that our denoising method, based on the analytic properties of
the
\textit{Z-transform} (generating function) of the data \cite{zeta}, and its
subsequent J-Matrix we associate to it, promises a significantly improvement
of the  detection probability even in the case of highly correlated noise.

Spectral analysis has been dominated by the Fourier transform (and its
variants), which is the ideal tool when noise is not too important.

The major drawback of the Fourier transform is its \textit{linearity} which
prevents it to distinguish noise from signal because the Fourier Transform
treats the noise on the same foot as the signal, and therefore leaves noise
intact.

\textit{Furthermore it is clear that the discrete Fourier Transform is
nothing but the specialization of the Z-transform to specific points of the complex plane which are the roots of
unity.} This is important to notice because, as we have shown in Ref. \cite{jpa},
\textit{the roots of unity are noise attractors in the complex plane. }That
is to say, noise is \textit{not} distributed uniformly in the complex plane
and one should take advantage of this fact to build an algorithm that makes
use of this knowledge.

\textit{This also explains the poor results of data spectral analysis in the
presence of heavy noise produced by the FFT.}

One of the most important application of the new introduced \textit{J-transform} concerns damped oscillating
signals, such as those found in nuclear magnetic resonance data \cite{bel} and in burst gravitational waves
\cite{bur,ring}.

The present paper is thus organized:

We first show how to construct the successive J-Matrix approximations built
on the time-series data.

We then summarize the principal properties of said approximations and
outline how such properties can be used to overcome current detection
limitations.

Finally, we give an example where we simulate a search for Gravitational Waves Bursts: the aim is to detect a weak
short signal embedded in a single sequence of highly correlated noise (very long, compared to the signal length)
and to determine the time at which it happens. Our provisional procedure succeeds to detect the signal in a
simulated time sequence having a rather poor signal to noise ratio.

\section{The tridiagonal Jacobi Matrix.}\label{s2}

Given a time-series $s_{0},s_{1},s_{2},.....s_{2n},.......$, it is a standard practice \cite{zeta} to associate to
it its \textit{Z-transform} defined as:

\begin{equation}
Z(z)=\sum_{m\geq 0}s_{m}z^{-m}  \label{1}
\end{equation}

It has been shown \cite{jpa} that the poles of the $\left[ \frac{n-1}{n}\right] (z)$ Pad\'{e} Approximant to
$Z-transform$ are the eigenvalues of the tridiagonal $n\times n$ Jacobi Matrix

\begin{equation}
J_{n}=\begin{bmatrix} A_{0} & 1 & 0 & ... & 0 & 0 \\ R_{1} & A_{1} & 1 & 0 &
0 & 0 \\ 0 & R_{2} & A_{2} & 1 & ... & 0 \\ ... & & & & 1 & 0 \\ 0 & ... & 0
& R_{n-1} & A_{n-1} & 1 \\ 0 & 0 & ... & 0 & R_{n} & A_{n}\end{bmatrix}
\end{equation}

where

\begin{equation}
A_{k}=-(a_{2k}+a_{2k+1})\qquad R_{k}=a_{2k-1}a_{2k}\qquad k\geq 1\qquad
a_{0}=0
\end{equation}

The $a_{k}$ being the coefficients of the Stieltjes continued fraction
expansion of the $Z-transform$.

The zeros of the Z- transform Pad\'{e} Approximant can be calculated in a similar way, diagonalizing the $(n-1)
\times (n-1)$ Jacobi Matrix for the numerator of the Pad\'{e} Approximant, which is simply the denominator Jacobi
Matrix with the first column and first row removed.

The procedure for computing both zeros and poles of the Z-transform
therefore consists in the diagonalization of an already tridiagonal matrix;
this makes it computationally extremely efficient (the Hessenberg matrix,
often used to find the roots of a polynomial is only sparse and therefore
less efficient), and \textit{allows us to easily calculate hundreds of poles
and zeros.}

\subsection{The spectrum of the Jacobi Matrix.}

To any bounded noisy time series can be associated a tridiagonal Hilbert space operator, its $J-Operator$,
extension when $n\rightarrow \infty$ of the previously defined $J_{n}$. With probability one (that is: except for a
set of input data series of measure zero), the spectrum of this operator is made of two parts ( see Ref. \cite{jpa}
and references therein):

1) A discrete spectrum, made of a finite number of eigenvalues inside the
unit circle: each eigenvalue represents a component of the signal made of a
finite number of damped oscillators. The corresponding eigenfunctions have
\textit{finite norm}.

From the value $z_{p}$ of the $p^{th}$ eigenvalue, the characteristics of
the damped oscillators of the signal can be immediately derived: the
frequencies
\begin{equation}
f_{p}=\frac{2n}{T}\frac{\arg z_{p}}{2\pi }\label{freq}
\end{equation}
and the damping factors
\begin{equation}
\alpha _{p}=\frac{2n}{T}\log {|z_{p}|}
\end{equation}

where $2n$ is the number of data, and $T$ the recording time. ($\frac{2n}{T}$ is the sampling rate)

2) An essential spectrum with support the unit circle. This spectrum is
associated with the noise (uncorrelated part of the signal). The
corresponding eigenfunctions have \textit{infinite norm}.

\begin{figure}[htbp]
\centering\epsfig{file=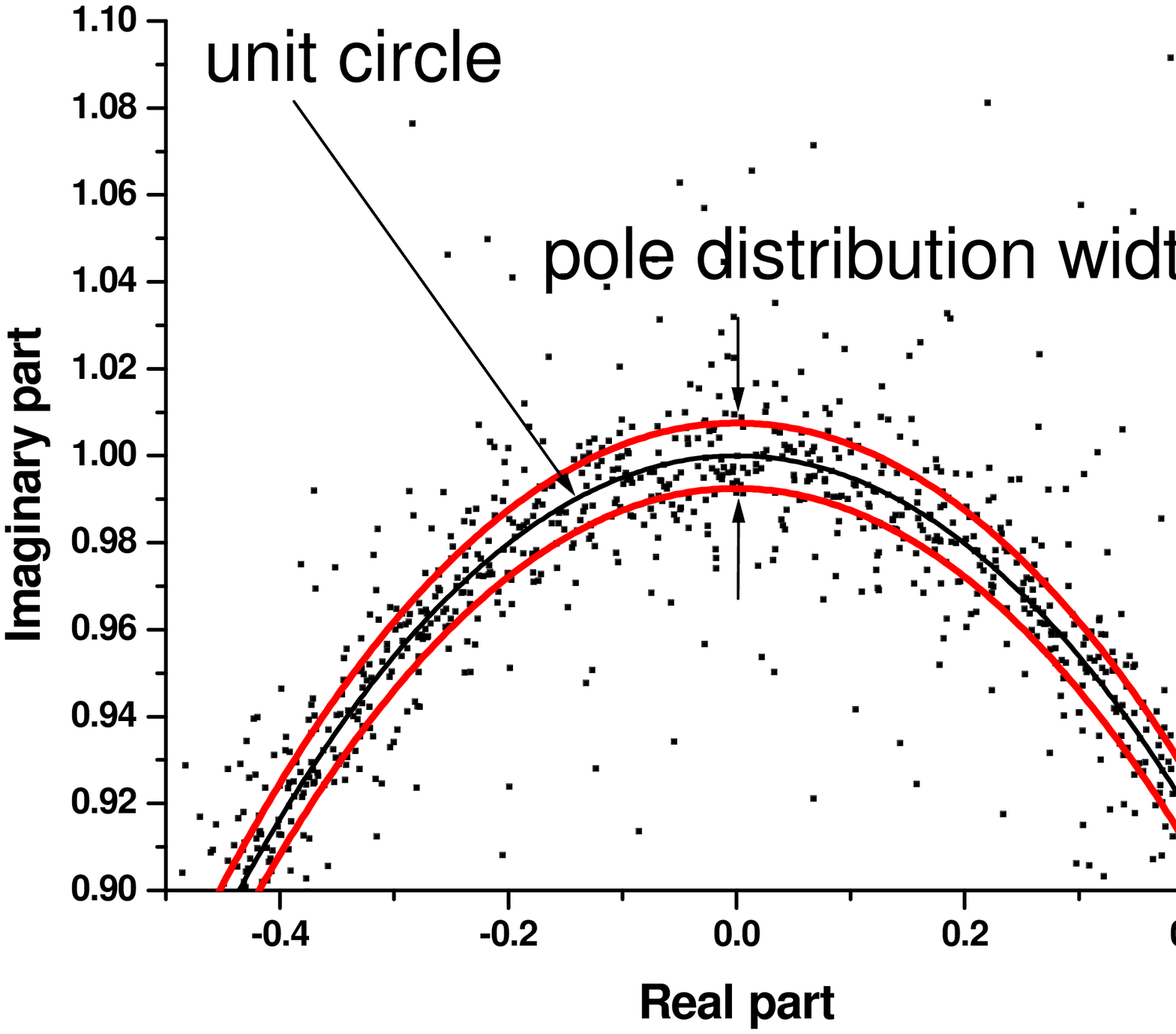,width=0.7\linewidth} \vspace{-0.3in}
\vspace{-0.4in}
\caption{An example of distribution of the noise poles around the unit
circle: shown are the poles for $40$ $[149/150]$ Pad\'{e} Approximants. The
width of the distribution is independent from the noise amplitude, but it
depends on the order of the Pad\'{e} Approximant.}
\label{fig1}
\end{figure}

At a finite order, when the infinite $J-Operator$ is truncated, it decomposes into the Froissart poles of the
Froissart doublets \cite{dou1,dou2,dou3,dou4} (see Figure (\ref{fig1})). The distribution of the poles and zeros of
these doublets is universal \cite{jpa}: independently from the kind of noise, the radial distribution is Lorentzian
around the unit circle, with width proportional to$\frac{1}{2n}\ln {2n}$, where $2n$ is the number of data points.
The phase distribution is uniform (approaching the $n$ roots of unity when $2n$ goes to infinity).

\begin{figure}[htbp]
\centering\epsfig{file=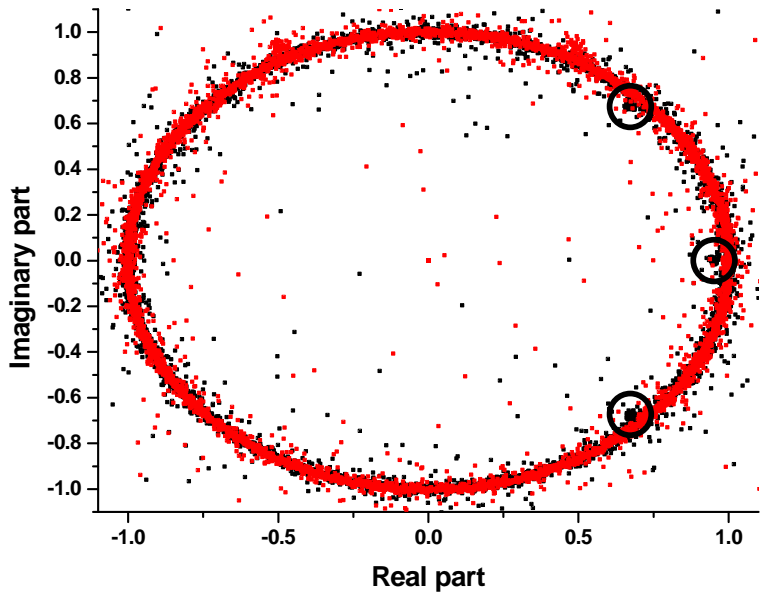,width=0.7\linewidth} \vspace{-0.3in}
\vspace{-0.4in}
\caption{Poles (black) and zeros (red) for $40$ noisy time series of $300$
points, each containing the same damped nonoscillating signal, an
oscillating one with the same damping factor (again equal for all runs), and
a different realization of gaussian noise. The circles indicate the
positions of the signal poles.}
\label{fig2}
\end{figure}

We start by giving a simple straightforward image of the feature of the method. Figure (\ref{fig2}) shows poles
(black) and zeros (red) of the $Z-transform$ for $40$ different noisy time series of $300$ points each. Each
containing the same damped non-oscillating signal, an oscillating one with the same damping factor (again equal for
all runs), and a different realization of Gaussian noise. The matched filtering signal to noise ratio
$SNR={\frac{\sqrt{\Sigma _{p=1}^{2n}h_{p}^{2}}}{\sigma }}$, where $2n$ is the number of data points, $h_{p}$ are
the pure signal points and $\sigma $ is the noise variance, is in this case quite high: $30$ for the
non-oscillating signal and $15$ for the oscillating one, so as to have a very clear picture. Most of the poles are
around the unit circle and are covered by the corresponding zeros of the Froissart doublets; three clusters of
poles (circled in the Figure) inside the unit circle are instead clearly visible: their position is that of the
three signal poles.

Here, in order to introduce the method and for the convenience of the
reader, \textit{we have not yet made use of any of the sophisticated methods
developed in the sequel. That is the reason for the use of a high signal to
noise ratio.}

\section{Time reversal}

If one applies time reversal to a given time-series, one gets a new
time-series in which the last term has become the first one and vice-versa.

It is not difficult to check that the denominator of the $Z-transform$ Pad\'{e} Approximant is \textit{covariant}
under time reversal.

In fact, calling $Q_{n}(z)$ the degree-n polynomial denominator of the Pad\'{e} Approximant \cite{pade}

\begin{equation}
Q_{n}=\begin{bmatrix} s_{0} & s_{1} & ... & s_{n} \\ s_{1} & s_{2} & ... &
s_{n+1} \\ ... & ... & ... & ... \\ s_{n-1} & s_{n} & ... & s_{2n-1} \\
z^{-n} & z^{-(n-1)} & ... & 1\end{bmatrix}.
\end{equation}

we have the covariant transformation

\begin{equation}
Q_{n}^{T}(z)=z^{n}Q_{n}(\frac{1}{z})
\end{equation}

where $Q_{n}^{T}(z)$ is the degree-n  polynomial denominator of the  Pad\'{e}
Approximant to the time reversal time-series.

Note that this is \textit{not} true for the zeros of the numerator of the Pad\'{e} Approximants to the
$Z-transform$.

Also, a less formal approach can be provided. The Z-transform of an infinite
time series is analytic outside the unit disk under the assumption that it
is bounded; this excludes the class of exponentially growing signals.

For a finite time series, exponentially growing signals are instead possible
and thus also poles outside the unit disk. From the fact that time inversion
transforms a damped signal in an exponentially growing one and vice versa,
it is clear that time inversion has to transform all poles $z_{p}$ into
their reciprocals $1/z_{p}$.

\section{Methods for extracting Spectra from Noisy Data.}\label{s4}

In order to make a clear distinction among the poles of the $Z-transform$ of a noisy time series, and identify the
eigenvalues corresponding to the discrete spectrum of the$J-Matrix$, we can make use of several properties of the
$Z-transform$. Each method we are going to present addressing a different property, a comparison of the results of
all methods applicable to any specific time-series is likely to be the optimal choice.

\subsection{Cleaning of Froissart doublets}

A first denoising method consists in removing from the complex plane the
poles which can be identified as part of Froissart doublets, due to their
proximity to zeros of the Z-transform.

This can be done by ordering the poles and zeros in couples in order of
increasing distance between the two, and keeping only those whose distance
is higher than a given value $\epsilon$ (usually $\epsilon \ge 0.2$).

\subsection{Variational principle}

We can on the other hand, making use of the results of section \ref{s2},
search the eigenvalues of Jacobi Matrix associated to the denominator of the
Z-transform having \textit{modulus strictly smaller than one} and for which
the norm of the eigenvector have local minima. Such eigenvalues correspond
to the frequencies of the signal.

\subsection{Stationariety}

It is also possible to compare Pad\'{e} approximants of different order and
look for stable ``non-zero paired" poles inside the unit circle: these will
be the signal poles, while the non stationary poles will be linked to the
noise. This method is particularly useful in dealing with poles in the tails
of the noise pole distribution.

\subsection{Undersampling}

Another method makes use of the fact that Pad\'{e} Approximations allow lossless undersampling for the class of
damped oscillators \cite{jpa}.

The above result suggests another technique -which we call ``interlaced sampling"- to improve sensitivity when only
a single data time-sequence is available. Taking advantage of the undersampling properties of the $Z-transform$, we
divide the data in $m$ undersampled sequences, the first one comprising the points $1,m+1,2m+1,3m+1,...$, the
second one comprising the points $2,m+2,2m+2,3m+2,...$ and and so on. To first order, the ratio of the width of the
pole distribution around the unit circle to the distance of the signal pole from it does not change, but instead of
a single signal pole, we do now get a cluster of $m$ poles, which are much easier to detect.

We point out here that we assume to be dealing with data series such that
the periods of the signal oscillations to be reconstructed cover several
data points. This means that, as long as $m$ is less than the number of data
points per period of each of the signal oscillators, the reconstructed
signal frequencies will be given by the $m^{th}$ roots of the poles $z_{p}$
having the lowest phases.

\subsection{Time reversal}

We have seen that time reversal transforms all poles into their reciprocals.
This suggests another way to increase the number of data poles: namely,
applying the same procedure using the data in both direct and inverted order.

As the coefficients of the Stieltjes continued fraction expansion of the $Z-transform$ are different in the two
cases, the numerically calculated poles will be reciprocal of each others in a nontrivial way.

\section{An example: Gravitational Waves Bursts}

\begin{figure}[htbp]
\centering\epsfig{file=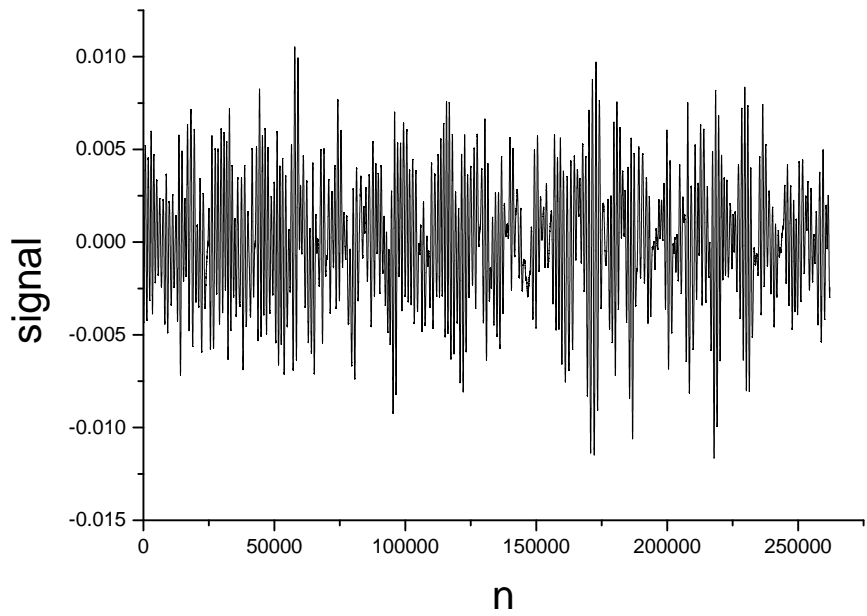,width=0.7\linewidth} \vspace{-0.3in}
\vspace{-0.4in}
\caption{The simulated interferometric data provided by R. Grosso. Here and
in the following figures, time is given in number of sampling periods, the
sampling rate being $16384Hz$.}
\label{fig3}
\end{figure}

\begin{figure}[htbp]
\centering\epsfig{file=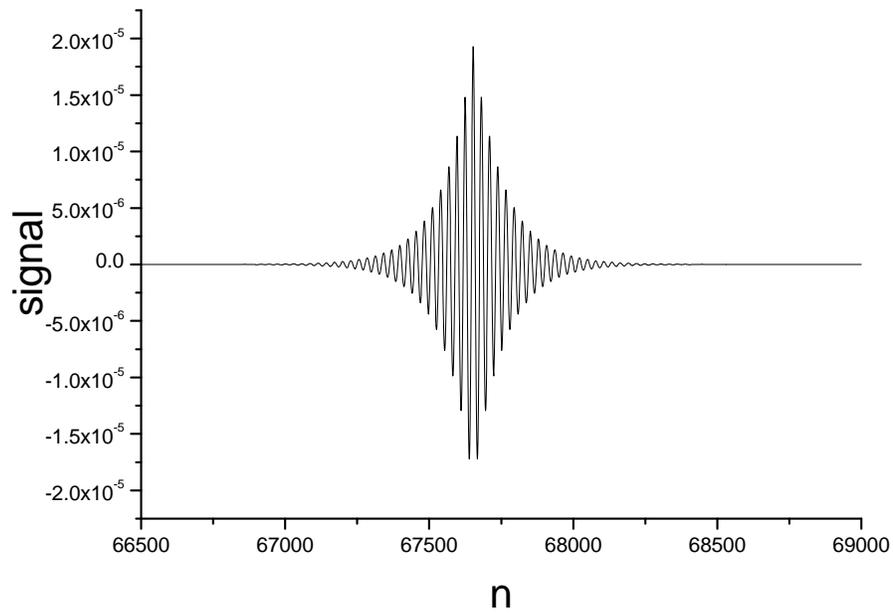,width=0.7\linewidth} \vspace{-0.3in}
\vspace{-0.4in}
\caption{The ``burst" injected in the data.}
\label{fig4}
\end{figure}

\begin{figure}[htbp]
\centering\epsfig{file=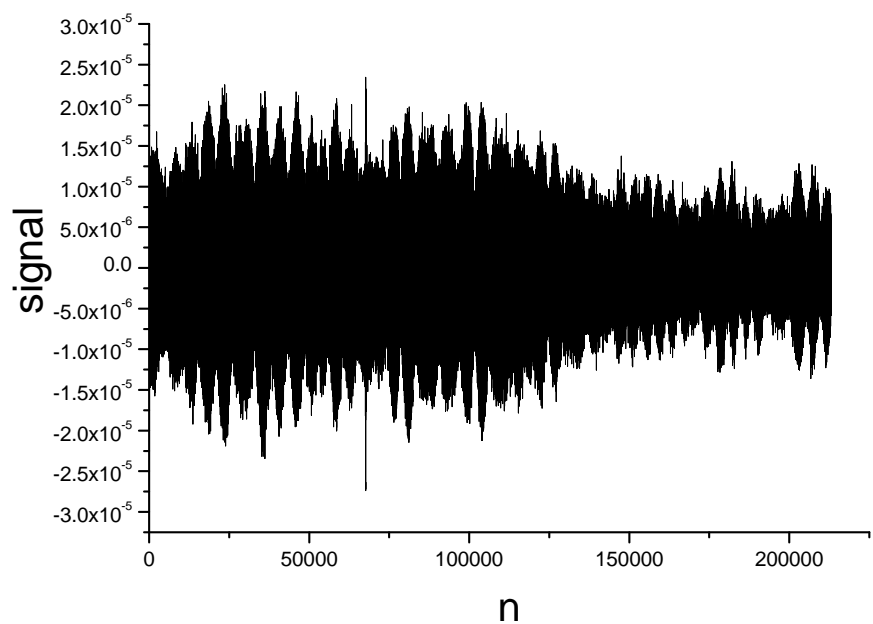,width=0.7\linewidth} \vspace{-0.3in}
\vspace{-0.4in}
\caption{The simulated data after being high-pass filtered at $200Hz$.}
\label{fig5}
\end{figure}

R.Grosso has provided us with simulated data (see Figure (\ref{fig3})) where a ``burst" (see Figure (\ref{fig4}))is
injected on a background ``noise" accurately reproducing the correlations found in the data from gravitational
waves interferometric detectors \cite{grosso}; the matched filtering signal to noise ratio is $SNR \simeq 0.04$.
Once the frequencies below $200Hz$, known to contain no signal in the present case, have been filtered out, the
``burst" peak amplitude is approximately equal to the background amplitude (see Figure (\ref{fig5})) and the $SNR$
grows to $24$. {\it No assumpion is made on the signal, except that it's a burst whose duration is of the order of
$0.1s$ and whose frequency is higher than $200Hz$}.

The basic problem, consisting in detecting a short signal embedded in a
single very long sequence and the time at which it happens, is complicated
by the fact that the ``noise" includes strong fake signals and is therefore
highly correlated; this results in a non uniform distribution of the poles
and zeros on the unit circle.

Several techniques are currently being used to deal with these two problems
in a way that requires direct reading by the operator of only a small set of
data, such as change point detection and time-frequency methods. Examples of
the former are the BlockNormal method \cite{bno}, the Excess Power Search
method \cite{epsm}, and the ALF (alternative linear fit) filter \cite{alf}.
Examples of the latter are The WaveBurst algorithm \cite{wb} and the PSD
(power spectral density) method \cite{psd}.

Up to now, the minimum matched filtering signal to noise ratio allowing signal detection at a significant level
using the above methods is around $snr \simeq 15$, assuming Gaussian noise; and $snr \simeq 8$ when the signal is
known and it's possible to use matched filtering \cite{sumy}. It is, on the other hand, much higher when the above
two conditions are not met \cite{sumy}, as in the present case.

The provisional procedure we describe in the following can be viewed as an extension to the complex plane of the
time-frequency methods; it is in no way definitive. Our aim here is to show that the J-transform can be an
effective tool in this kind of search:

1) We divide the sequence in 235 blocks of 903 data each.

2) The data in each block gets divided in 3 undersampled sequences of 301
data each, taking the points 1,4,7,10,13,16,19... for the first sequence,
the points 2,5,8,11,14,17,20... for the second one and 3,6,9,12,15,18,21...
for the third one (``interlaced sampling").

3) For each undersampled sequence, we clean the Froissart doublets: we
calculate the poles and zeros of the [149/150] Pad\'{e} approximant, we
order the poles and zeros in couples in order of increasing distance between
the two and keep only those whose distance is higher than 0.2.

4) For each block, we divide the complex plane close to the unit circle in
boxes (either square ones or by radius and angle) and take the difference
between the number of poles and the number of zeros in each of the boxes.

5) The 3 sequences of each block essentially cover the same signal sequence,
they should therefore all 3 give an isolated pole if there is a signal; to
be on the safe side we accept 2 out of 3 and therefore keep only the boxes
where said number is strictly higher than 1. To make sure not to miss
multiple isolated pole occurrences close to the box sides, we repeat the
procedure shifting the boxes by 1/2 step.

\begin{figure}[htbp]
\centering\epsfig{file=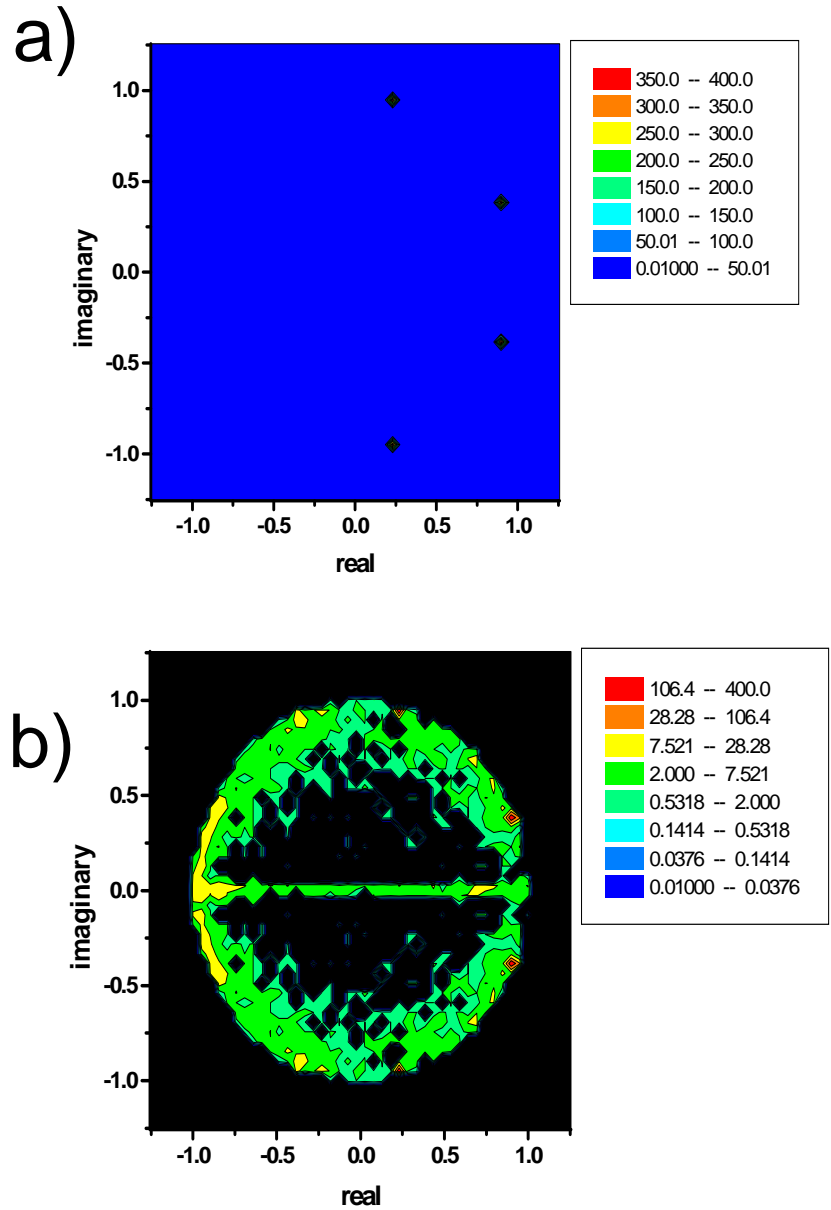,width=0.7\linewidth} \vspace{-0.3in}
\caption{The number of times isolated poles are observed at any given
location of the complex plane. a) linear spacing of the level curves. b) logaritmic spacing of the level curves.
The four most noticeable peaks are at $\protect\rho = 0.97324$, $\protect\varphi = \pm 1.33232$ and at
$\protect\rho = 0.98766$, $\protect\varphi = \pm 0.40439$.}
\label{fig7}
\end{figure}

6) Due to the non uniform distribution of the poles and zeros on the unit
circle, we observe a number of boxes selected at point 5) which appear in
many blocks (see Figure (\ref{fig7})). To get rid of them, we count the
number of blocks for which the result for a given box is strictly higher
than 1. Boxes for which this number is higher than 3 are discarded.

7) Of the remaining boxes, we keep those appearing only once, or more times
but only if this happens in consecutive blocks.

We are left with no consecutive events and only $9$ non-repeated events,
each with two poles; closer inspection shows for blocks 63, 76, 215 a third
isolated pole close to the two found. (actually, since the data are real, a
couple of complex conjugate poles near the other two couples). In all cases
it is nonetheless too far away to be reasonably considered related.

8) Considerations of two kinds allow to drop all but block 76. Namely the poles are extremely close to the unit
circle, resulting in a signal which is almost constant in amplitude and therefore not a burst, and/or they appear
in boxes contiguous to those corresponding to strong noise peaks in Figure (\ref{fig7}).

\begin{figure}[htbp]
\centering\epsfig{file=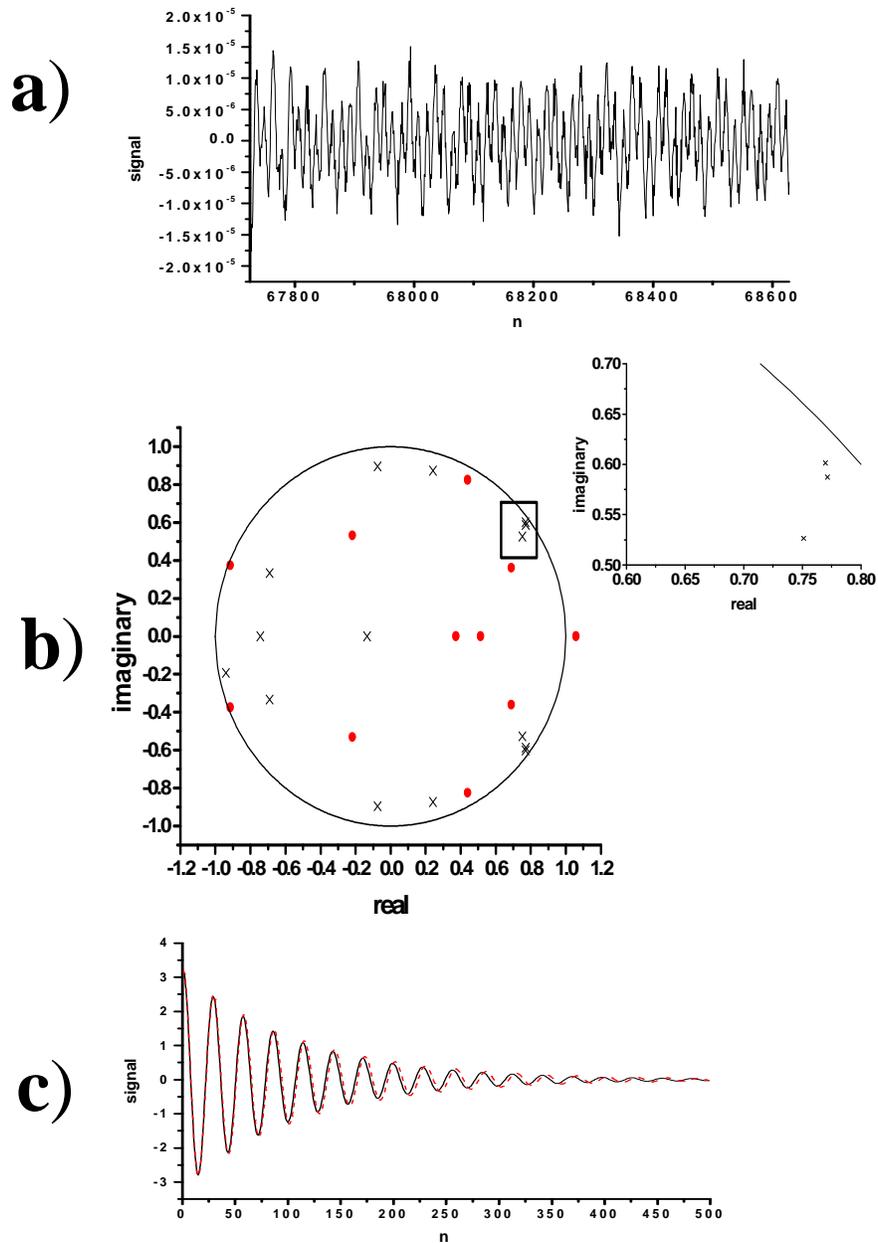,width=0.7\linewidth} %\vspace{-0.9in}
\caption{a) The time sequence for block 76 of the simulated ``burst" data
produced by R. Grosso (signal plus noise). b)Poles (black crosses) and zeros (red circles) selected in step 3) of
the procedure; to guide the eye, the unit circle is drawn in black. The highlighted square is magnified in the
inset and shows the two poles in the upper complex planes individuated in step 7) of the procedure as corresponding
to the signal (the complex conjugate poles in the lower complex plane can be seen in the full picture). c) the
reconstructed signal (red dash) compared to the original one injected into the noise (black full line). The
reconstructed signal is obtained using (\ref{freq}) with $\protect\rho = 0.97331$ and $\protect\varphi = 0.65711$
given by the average of the positions of the two poles shown in Figure
\protect\ref{fig8}b above. The vertical scale is arbitrary.}
\label{fig8}
\end{figure}

The average position of the selected poles for block 76 allows to
reconstruct the clean signal, as shown in Figure (\ref{fig8}). As we have
not performed a search for the exact position of the signal peak, the signal
amplitude was not calculated.

Even using ``interlaced sampling" we only have three cases per block. To have some idea of the order of detection
probability and false alarm rate, we repeated the search shifting the blocks by $1/4$, $1/2$, and $3/4$ of their
length. Considering all $4$ runs, at the end of step 7) only two events appear more than once: the injected signal
(3 times out of 4: shift $0$, $1/2$, and $3/4$), and a second event, close to the unit circle, appearing at shifts
$1/4$ and $1/2$.

More extensive tests will be needed to properly determine both detection
probability and false alarm rate, but the result is encouraging.

\section{Conclusion.}

In our method, the Z-transform appears as an extension of the discrete Fourier transform to complex values of the
frequency: in the complex plane the discrete Fourier transform is the restriction of the Z transform to values of z
located at the roots of unity. However, the noise is not uniformly distributed in the complex z-plane and,
unfortunately, the roots of unity are noise attractors as shown in Ref. \cite{jpa}. Therefore, the Fourier
transform is not a good choice when a damped signal is embedded in high noise as it is the case in many
circumstances. The analytic treatment of the noise we propose, distinguishes, in a drastic way, the signal from the
noise by their totally different analytic properties. The fact that the noise presents an analytic characterization
that is universal, independent of its statistical properties, has as consequence a much greater independence of the
signal identification on the noise level.

The example we have given strongly suggests that the $J-transform$ analysis
-especially if coupled with judicious undersampling- promises \textit{not
only in theory} to be an extremely powerful tool for the analysis of
interferometric gravitational wave burst data.

There is still plenty of room for improvement: of the methods outlined in section \ref{s4} we have used, in the
example given above, only the cleaning of Froissart doublets, the knowledge that the signal poles must be inside
the unit circle, and ``interlaced sampling".

We are moreover still not using all the information available. For example
we did not use the residues of the poles, from which the signal amplitude
can be obtained, and we haven't yet performed a complete optimization over
the various parameters used in the analysis. In particular there is the need
for an extensive mapping of the spread of the reconstructed poles as a
function of noise level, damping factor, and Pad\'{e} approximant order:
block length, number of interlaced samples and cutting length would all be
affected by this.

\section{Acknowledgments}

We thank Professor Roberto Grosso from the Universitaet Erlangen-Nuernberg,
for building for us mock noisy gravitational waves data

We thank Professor Marcel Froissart, from College de France, for discussions
and suggestions.

We thank Professor Carlos Handy, Head of the Physics Department at Texas
Southern University, for his support.

We thank Professor Bernhard Beckermann of Lille university, France for
explaining to us his method of computing Pad\'{e} Approximations avoiding
small denominators instabilities.

Special thanks to Professor Mario Diaz, Director of the Center for
Gravitational Waves at the University of Texas at Brownsville: without his
constant support this work would have never been possible.

\bigskip

Supported by sub-award CREST to the center for Gravitational Waves, Texas
University at Brownsville, Texas USA

\bigskip

%\begin{references}

%\end{references}

\newpage .

\vspace{0.2in}

\vspace{0.9in}

\newpage .

\vspace{0.4in}

\vspace{0.9in}

\newpage .

\vspace{0.9in}

\vspace{0.2in}

\newpage

\vspace{0.2in}

\end{document}